\documentstyle[preprint,aps]{revtex}

\def\be{\begin{equation}}
\def\en{\end{equation}}
\def\bq{\begin{eqnarray}}
\def\eq{\end{eqnarray}}

\begin{document}

{\large On the uniqueness of the surface sources of evoked potentials }$\ $%
%
%
%
%
%
%
%
%

\begin{center}
{\bf \vspace{1in}}

Alejandro Cabo$^{*}$, Carlos Handy\vspace{0.25in} and Daniel Bessis.

{\it Center of Theoretical Studies of Physical Systems}

{\it Clark Atlanta University}, {\it Atlanta U.S.A.\vspace{0.1cm}\bigskip 
\vspace{1in}}

{\bf Abstract.}

\noindent
The uniqueness of a surface density of sources localized inside a spatial
region $R$ and producing a given electric potential distribution in its
boundary $B_0$ is revisited. The situation in which $R$ is filled with
various metallic subregions, each one having a definite constant value for
the electric conductivity is considered. It is argued that the knowledge of
the potential in all $B_0$ fully determines the surface density of sources
over a wide class of surfaces supporting them. The class can be defined as a
union of an arbitrary but finite number of open or closed surfaces. The only
restriction upon them is that no one of the closed surfaces contains inside
it another (nesting) of the closed or open surfaces.

\strut 

\vspace{1in}

* On leave in absence from the: {\it \ Group of Theoretical Physics,
Instituto de Cibern\'{e}tica, Matem\'{a}tica y F\'{i}sica, Calle E,} {\it %
No. 309 Vedado, La Habana, Cuba.}

{\it email: cabo@cidet}.{\it icmf.inf.cu}

\newpage
\end{center}

\section{Introduction.}

The uniqueness problem for the sources of the evoked potential in the brain
is a relevant research question due to its role in the development of
cerebral electric tomography\cite{amir}, \cite{riera} , \cite{nunez}, \cite
{aine} . Since long time ago, it is known that the general inverse problem
of the determination of volumetric sources from the measurement of the
potential at a surface is not solvable in general\cite{hemholtz},\cite
{hadamard}. However, under additional assumptions about the nature of the
sources, solutions can be obtained \cite{scherg},\cite{wahba},\cite{riera1}.
The supplementary assumptions can be classified in two groups: the
physically grounded ones, which are fixed by the nature of the physical
problem and the ones which are imposed by invoking their mathematical
property of determining a solution, but having in another hand, a weak
physical foundation. The resumed situation implies that the determination of
physical conditions implying the uniqueness of the sources for the evoked
potentials remains being an important subject of study. Results in this
direction could avoid the imposition of artificial conditions altering the
real information on the sources to be measured.

The question to be considered in this work is the uniqueness of the sources
for evoked potentials under the assumption that these sources are localized
over surfaces. This issue was also treated in Ref. [1] by including also
some specially defined volumetric sources. The concrete aim here is to
present a derivation of the results enunciated in [1] for the case of open
surfaces and to generalize it for a wider set of surfaces including closed
ones.

We consider that the results enunciated in Ref. [1] are valid and useful
ones. Even more, we think that a relevant merit of that paper is to call for
the attention to the possibility for the uniqueness for classes of surface
density of sources. Specifically, in our view, the conclusion stated there
about the uniqueness of the sources of evoked potentials as restricted to
sources distributed in open surfaces is effectively valid. In the present
work, the central aim is to extend the result for a wider set of surfaces
including closed ones by also furnishing an alternative way to derive the
uniqueness result. The uniqueness problem for the special class of
volumetric sources discussed in [1] is not considered here in any way.

The physical system under consideration is conformed by various volumetric
regions, each of them having a constant value of the conductivity, separated
by surface boundaries at which the continuity equations for the electric
current is obeyed. It should pointed out that the special volumetric sources
examined in Ref. [1] are not addressed here. The precise definition of the
generators under examination is the following. The sources are assumed to be
defined by continuous and smooth surface densities lying over a arbitrary
but finite number of smooth open or closed surfaces. The unique constraint
to be imposed on these surfaces is that there is no nesting among them. That
is, there is no closed surface at which interior another open or closed of
the surfaces resides. This class of supports expands the one considered in
Ref. [1] and in our view is sufficiently general to create the expectative
for the practical applications of the results. It should be stressed that
the boundaries between the interior metallic regions are not restricted by
the ''non-nesting'' condition. That is, the fact that the skull and the few
boundaries between cerebral tissues can be visualized as nearly closed
surface does not pose any limitation on the conclusion. The ''non-nesting''
condition should be valid only for the surfaces in which the sources can be
expected to reside. For example, if by any mean we are sure that the sources
stay at the cortex surface, then the uniqueness result apply whenever the
portion of the cortex implied does not contains any closed surface.

The paper is organized as follows. An auxiliary property is derived in the
form of a theorem in the Section II. In Section III the proof of uniqueness
for the kind of sources defined above is presented.


\section{Green Theorem and Field Vanishing Conditions}

Let us consider the potential $\phi $ generated by a source distribution
concentrated in the ''non-nested'' set of open or closed surfaces defined in
last Section, which at the same time are contained within a compact and
simply connected spatial region $R.\ $The set $R,$ as explained before, is
formed by various connected subregions $R_i,\ i=0,1,...n$ each of them
filled with a metal having a constant conductivity $\sigma _i$. Also, let $%
B_{ij}$ the possibly but non necessarily existing, boundary between the
subregions $R_i$ and $R_j$ and $B_0\ $the boundary of $R.$ For the sake of a
physical picture, we can interpret $B_0$ as the surface of the skull, $R$ as
the interior of the head and the subregions $R_i$ as the ones containing the
various tissues within the brain. It is defined that the exterior space of
the head corresponds to $R_0$. In addition, let $S_i,\ i=1,...m\ $ the
surfaces pertaining to the arbitrary but finite set $S$ of non-nested open
or closed surfaces in which the sources are assumed to be localized. The
above mentioned definitions are illustrated in Fig.1.

Then, the Poisson equation satisfied by the potential $\phi $ in the
interior region of $R$ can be written as

\begin{eqnarray}
\nabla ^2\phi \left( \overrightarrow{x}\right) &=&\frac{g\left( 
\overrightarrow{x}\right) }{\sigma \left( \overrightarrow{x}\right) }, \\
g\left( \overrightarrow{x}\right) &=&-\overrightarrow{\nabla }.%
\overrightarrow{J}\left( \overrightarrow{x}\right) ,
\end{eqnarray}
where$\overrightarrow{J}$ are the impressed currents (for example, generated
by the neuron firings within the brain) and the space dependent conductivity
is defined by

\begin{equation}
\sigma \left( \overrightarrow{x}\right) =\sigma _i\ \ \ for\ \overrightarrow{%
x}\in R_i.
\end{equation}

It should be noticed that the conductivities are different from zero only
for the internal regions to $R.$ The vacuum outside is assumed to have zero
conductivity and the field satisfying the Laplace equation. In addition
outside the support of the sources where $g=0$ the Laplace equation is also
satisfied.

The usual boundary conditions within the static approximation, associated to
the continuity of the electric current at the boundaries, take the form

\begin{equation}
\sigma _i\frac{\partial \phi }{\partial n_i}\mid _{x\ \in \ B_{ij}}=\sigma _j%
\frac{\partial \phi }{\partial n_j}\mid _{x\ \in \ B_{ij}},
\end{equation}
where $\partial n_i$ symbolizes the directional derivative along a line
normal to $B_{ij}$ but taken in the limit of $x->B_{ij}\ $from the side of
the region $R_i.$

A main property is employed in this work in obtaining the claimed result. In
the form of a theorem for a more precise statement it is expressed as

{\bf Theorem.}

Let $\phi $ is a solution of the Laplace equation within an open and
connected spatial region $R^{*}$. Assume that $\varphi \ $have a vanishing
electric field over an open section of certain smooth surface $S^{*}$ which
is contained in an open subset $Q$ of $R^{*}$. Let the points of the
boundaries between $Q$ and $R^{*}$ have a minimal but finite distance among
them. Then, the potential $\phi $ is a constant over any open set contained
in $R^{*}.$

As a first stage in the derivation of this property, let us write the Green
Theorem as applied to the interior of the open region $Q\ $defined in the
Theorem 1 in which a field $\varphi $ satisfies the Laplace equation. Then,
the Green Theorem expresses $\varphi $ evaluated at a particular interior
point $\overrightarrow{x}$ in terms of itself and its derivatives at the
boundary $B_Q$ as follows.

\begin{equation}
\varphi \left( \overrightarrow{x}\right) =\int_{B_Q}d\overrightarrow{%
s^{^{\prime }}}.\left( \frac 1{\left| \overrightarrow{x}-\overrightarrow{%
x^{^{\prime }}}\right| }\overrightarrow{\nabla _{x^{^{\prime }}}}\ \varphi
\left( \overrightarrow{x^{^{\prime }}}\right) -\overrightarrow{\nabla
_{x^{^{\prime }}}}\left( \frac 1{\left| \overrightarrow{x}-\overrightarrow{%
x^{^{\prime }}}\right| }\right) \varphi \left( \overrightarrow{x^{^{\prime }}%
}\right) \right)
\end{equation}
where the integral is running over the boundary surface $B_Q$ which is
described by the coordinates $\overrightarrow{x^{^{\prime }}}.$ This
relation expresses the potential as a sum of surface integrals of the
continuous and bounded values of $\varphi $ and its derivatives. Those
quantities are in addition analytical in all the components of $%
\overrightarrow{x},$ if the point have a finite minimal distance to the
points in $B_Q.$ These properties follow because $Q$ $\subset R^{*}$ and
then, $\varphi $ satisfies the Laplace equation in any open set in which $Q$
and its boundary is included. But, due to the finite distance condition
among the point $\overrightarrow{x}$ and the points of $B_Q$, the expression
(5) for $\varphi $ should be an analytical function of all the coordinates
of $\overrightarrow{x}.$ Figure 2 depicts the main elements in the
formulation of the Green Theorem.

Further, let us consider that $S^{*}$ is siting inside the region $Q.$ Then,
as this surface is an equipotential and also the electric field over it
vanishes, it follows that no line of force can have a common point with it.
This is so because the divergence of the electric field vanishes, then it is
clear that the existence of nonvanishing value of the electric field at
another point of the line of force will then contradicts the assumed
vanishing of the divergence. Therefore, the lines of forces in any
sufficiently small open neighborhood containing a section of $\ S^{*}$
should tend to be parallel to this surface on approaching it, or on another
hand, the electric field should vanish. Next, it can be shown that in such
neighborhoods the lines of forces can not tend to be parallel.

Let us suppose that lines of forces exist and tend to be tangent to the
surface $S^{*}$ and consider the integral form of the irrotational property
of the electric field as

\begin{equation}
\oint_C\overrightarrow{E}.\ d\overrightarrow{l}=\int_{C_1}\overrightarrow{E}%
.\ d\overrightarrow{l}+\int_{C_2}\overrightarrow{E}.\ d\overrightarrow{l}%
=\int_{C_1}\overrightarrow{E}.\ d\overrightarrow{l}=0
\end{equation}
where the closed curve $C$ is constructed as follows: the piece $C_1\ $%
coincides with a line of force, the piece $C_2$ is fixed to rest within the
surface $S^{*}$ and the other two pieces necessary to close the curve are
selected as being normal to the assumed existing family of lines of forces.
The definitions are illustrated in Fig. 3. By construction, the electric
field is colinear with the tangent vector to $C_1$ and let us assume that we
select the segment of curve $C_1\ $for to have a sufficiently short but
finite length in order that the cosine associated to the scalar product will
have a definite sign$\ $in all $C_1$. This is always possible because the
field determined by (5) should be continuous. Then Eq. (6) implies that the
electric field vanish along all $C_1$ as a consequence of the integrand
having a definite sign and then should vanish identically. Since this
property is valid for any curve pertaining to a sufficiently small open
interval containing any particular open section of $S^{*},$ it follows that
in certain open set containing $S^{*}\ $there will be are no lines of
forces, or what is the same, the electric field vanish.

To finish the proof of the theorem, it follows to show that if $\varphi $
and the electric field vanish within a certain open neighborhood $N,$
included in an arbitrary open set $O$ pertaining to the region $R^{*}$ in
which the Laplace equation is obeyed, then $\varphi $ and the electric field
vanish in all $O$ . Consider first that $Q$ is an open set such that $%
O\subset Q$ and also suppose that the smallest distance form a point in $O$
to the boundary $B_Q$ of $Q$ has the finite value $\delta $. Then, the Green
Theorem (5) as applied to the region $Q$ expresses that the minimal radius
of convergence of $\varphi \ $considered as analytical function of any of
the coordinates is equal or greater than $\delta .$

Imagine now a curve $C$ starting in an interior point $P$ of $N$ and ending
at any point $P_1$ of $O.$ Assume that$\ C$ is formed by straight lines
pieces (See Fig. 4). It is then possible to define $\varphi $ as a function
of the length of arc $s$ of $C\ $ as measured form the point $P$. It should
be also valid that in any open segment of $C,$ not including the
intersection point of the straight lines, the potential $\varphi \ $is an
analytical function of $s.\ $Furthermore, let consider$\ C\ $as partitioned
in a finite number of segments of length $\sigma <\delta .$ Suppose also,
that the intersection points of the straight lines are the borders of some
of the segments. It can be noticed that $\varphi $ vanishes in any segment
of $C$ starting within $N$ because it$\ $vanishes$\ $in $N\ $exactly. Thus,
if $\varphi \ $and the electric field are not vanishing along all $C,\ $%
there should be a point over the curve in which the both quantities do not
vanish for an open region satisfying $s>s_o,$ and vanish exactly for another
open interval obeying $s<s_o$. However, in this case, all the derivatives of 
$\varphi $ of the electric field over $s$ vanish at $s_o.\ $This property in
addition with the fact that the Taylor series around $s_o$ should have a
finite radius of convergence $r>\delta ,$ as it assumed in the Theorem 1,
leads to the fact that $\varphi $ and the electric field should vanish also
for $s>s_o.$ Henceforth, the conclusion of the Theorem 1 follows: the
potential $\varphi $ and its corresponding electric field vanish at any
interior point of $R^{*}.$

\section{Uniqueness of the Non-Nesting surface sources}

Let us argue now the uniqueness of the sources which are defined over a set
of non nested surfaces $S\ $ producing specific values of the evoked
potential $\phi $ at the boundary $B_0\ $of the region $R.$ For this purpose
it will be assumed that two different source distributions produce the same
evoked potential over $B_0.$ The electrostatic fields in all space
associated to those sources should be different as functions defined in all
space. They will be called $\phi _{1\text{ }}$and $\phi _2.$ As usual in the
treatment of uniqueness problems in the linear Laplace equation, consider
the new solution defined by the difference $\phi =$ $\phi _{1-}\phi _2.$
Clearly $\varphi $ corresponds to sources given by the difference of the
ones associated to$\ \phi _1$ and $\phi _2.$ It is also evident that $\phi $
has vanishing values at $B_0.$ Then, since the sources are localized at the
interior of $R$ and $\phi $ satisfies the Laplace equation with zero
boundary condition at $B_0$ and at the infinity, it follows that the field
vanishes in all $R_0,\ $that is, in the free space outside the head.
Therefore, it follows that the potential and the electric field vanish in all%
$\ B_0\ $when approaching this boundary from the free space ($R_0).$ The
continuity of the potential, the boundary conditions (3) and the
irrotational character of the electric field allows to conclude that $\phi $
and the electric field also vanish at any point of $B_0$ but now when
approaching it from any interior subregion $R_i$ having a boundary $B_{i0}$
with the free space. Moreover, if the boundary surface of any of these
regions which are in contact with the boundary of $R$ is assumed to be
smooth, then it follows from Theorem 1 that the potential $\phi $ and its
the electric field vanish in all the open subsets of $R_i$ which points are
connected through its boundary $B_{i0}\ $with free space by curves
non-touching the surfaces of $S$. It is clear that this result hold for all
the open subsets of these $R_i$ in which Laplace equation is satisfied
excluding those which are also residing inside one of the closed surfaces $%
S_i$ in the set $S.$

It is useful for the following reasoning to remark that if we have any
boundary $B_{ij}$ between to regions $R_{i\text{ }}$ and $R_j,$ and the
potential $\phi $ and the electric field vanish in certain open (in the
sense of the surface) and smooth regions of it, then Theorem 1 implies that
the potential and the electric field also vanish in all the open subsets of $%
R_i$ and $R_j$ which are outside any of the closed surfaces in $S.$ Since
the sources stay at the surfaces in $S$ the field $\phi $ in some open
region of $R$ included inside certain of the closed surfaces $S_i$ will not
necessarily satisfy the Laplace equation in any interior point of $R$ and
Theorem 1 is not applicable.

Let us consider in what follows a point $P$ included in a definite open
vicinity of a subregion $R_i.$ Suppose also that $P$ is outside any of the
closed surfaces in $S$ . Imagine a curve $C\ $which join $P$ with the free
space and does not touch any of the surfaces in $S$. It is clear that, if
appropriately defined, $C$ should intersect a finite number of boundaries $%
B_{ij}$ including always a certain one $B_{j0}\ $with free space. Let us
also assume that $C\ $is adjusted in a way that in each boundary it crosses,
the intersection point is contained in a smooth and open vicinity (in the
sense of the surface) of the boundary (See Fig. 1 and 5). Then, it also
follows that the curve $C\ $can be included in open set $O_C$ having no
intersection with the non-nested surfaces in $S.$ This is so because the
region excluding the interior of the closed surfaces in $S$ is also
connected if the $S_i\ $are disjoint . But, from Theorem 1 it follows that $%
\phi \ $and the electric field must vanish in all $O_C.$ This should be the
outcome because the successive application of the Theorem 1 to the
boundaries intersected by the curve $C$ permits to recursively imply the
vanishing of $\phi \ $and the electric field in each of the intersections of 
$O_C$ with the subregions $R_i$ through which $C$ passes. The first step in
the recursion can be selected as the intersection of $C$ with $B_{j0}$ at a
point which by assumption is contained in an open neighborhood of the
boundary $B_{j0}$. As the electric field and $\phi $ vanish at free space,
the fields in the first of the considered intersection of $Oc\ $should
vanish. This fact permits to define another open and smooth neighborhood of
the next boundary intersected by $C\ $in which the field vanish and so on up
to the arrival to the intersection with the boundary of the region $R_i\ $%
containing the ending of $C\ $at the original point P. Therefore, the
electric field and the potential should vanish at an arbitrary point $P$ of $%
R$ with only two restrictions: 1) $P$ to be contained in an open
neighborhood of some $R_i$ and 2) $P$ to reside outside any of the surfaces
in $S.$ Thus, it is concluded that the difference solution $\phi $ and its
corresponding electric field, in all the space outside the region containing
the sources vanish. Henceforth, it implies that the difference between the
two source distributions also should be zero over any of the open surface in
the set $S.$ This is necessary because the flux going out from any small
piece of the considered surface is zero, which means that the assumed
continuous density of surface sources exactly vanish. This completes the
proof of the conclusion of Ref. [1] in connection with sources supported by
open surfaces. It only rests to show that the sources are also null over the
closed $S_i.$

Before continuing with the proof, it is illustrative to exemplify from a
physical point of view how the presence of nested surfaces among the $S_i$
destroys the uniqueness. For this aim let us let us consider that a closed
surface $S_i$ has another open or closed of the surface $S_j$ properly
contained inside it. That means that an open set containing $S_j$ is
contained inside $S_i.\ $Imagine also that $S_i$ is interpreted as the
surface of a metal shell connected to the ground; that is, to a zero
potential and that the surface $S_j$ is the support of an arbitrary density
of sources. As it is known from electrostatics theory, the charge density of
a metal connected to the ground is always capable to create a surface
density of charge at $S_i$ such that it exactly cancels the electric field
and the potential at the outside of $\ S_i,$ in spite of the high degree of
arbitrariness of the charge densities at the interior. That is, for nested
surfaces in $S$, it is not possible to conclude the uniqueness, because at
the interior of a nesting surface, and distributed over the nested ones,
arbitrary source distributions can exist which determine exactly the same
evoked potential at the outside boundary $B_0$.

Let us finally show that if no nesting exists the uniqueness also follows.
Consider any of the closed surfaces, let say $S_i.$ As argued before $\phi $
and the electric field vanish at any exterior point of $S_i$ pertaining to
certain open set containing $S_i.$ Then, the field created by the difference
between the sources associated to the two different solutions assumed to
exist should be different from zero only at the interior region. That zone,
in the most general situation can be filled by a finite number of metallic
bodies with different but constant conductivities. The necessary vanishing
of the interior field follows from the exact conservation of the lines of
forces for the ohmic electric current as expressed in integral form by

\begin{equation}
\int d\overrightarrow{s}.\ \sigma \left( \overrightarrow{x}\right) 
\overrightarrow{E}\left( \overrightarrow{x}\right) =0.
\end{equation}

Let us consider a surface $T\ $defined by the all the lines of forces of the
current vector passing through an arbitrarily small circumference $c$ which
sits on a plane being orthogonal to a particular line of force passing
through its center. Let the center be a point at the surface $S_i$ .
Because, the above defined construction, all the flux of the current passing
trough the piece of surface of $\ S_i\ ($which we will refer as $p)$
intersected by $T$ is exactly equal to the flux through any intersection of $%
T\ $with another surface determining in conjunction with $p$ a closed
region. By selecting a sufficiently small radius for the circumference $c$
it can be noticed that the sign of the electric field component along the
unit tangent vector to the central line of forces should be fixed. This is
so because on the other hand there will be an accumulation of charge in some
closed surface. Now, let us consider the fact that the electric field is
irrotational and examine a line of force of the current density which must
start at the surface $S_i.$ It should end also at $S_i$, because in another
hand the current density will not be divergence less. After using the
irrotational condition for the electric field in the form

\begin{equation}
\oint_C\overrightarrow{E}.\ d\overrightarrow{l}=\int_{C_1}\overrightarrow{E}%
.\ d\overrightarrow{l}+\int_{C_2}\overrightarrow{E}.\ d\overrightarrow{l}%
=\int_{C_1}\overrightarrow{E}.\ d\overrightarrow{l}=0
\end{equation}
in which $C_1$ is the line of force starting and ending at $S_i$ and $C_2$
is a curve joining the mentioned points at $S_i$ but with all its points
lying outside $S_i$ where $\phi =\phi _1$-$\phi _2\ $and the electric field
vanish. Let us notice that the electric field and the current have always
the same direction and sense as vectors, because the electric conductivity
is a positive scalar. In addition, as it is argued above, the current can
not reverse the sign of its component along the tangent vector of line of
forces. Therefore, it follows that also the electric field can`t revert the
sign of its component along a line of force. Thus, the integrand of the line
integral over the $C_1$ curve should have a definite sign at all the points,
hence implying that $\phi \ $and the electric field should vanish exactly in
all $C_1.$ Resuming, it follows that the electric field vanish also at the
interior of any of the closed surfaces $S_i.$ Therefore, the conclusion
arises that the difference solution $\phi =\phi _1$-$\phi _2=0$ in all the
space, thus showing that the evoked potential at $B_0$ uniquely fixes the
sources when they have their support in a set of non nesting surfaces $S.$

\noindent {\large {\bf Acknowledgments}}

We would like to thank the helpful discussions with Drs. Augusto
Gonz\'{a}lez , Jorge Riera and Pedro Vald\'{e}s. One of the authors ( A.C.)
also would like acknowledge the support for the development of this work
given by the Christopher Reynolds Foundation (New York,U.S.A.) and the
Center of Theoretical Studies of Physical Systems of the Clark Atlanta
University (Atlanta, U.S.A). The support of the Associateship Programme of
the Abdus Salam International Centre for Theoretical Physics (Trieste Italy)
is also greatly acknowledged. \newpage \newpage

{\large {\bf Figure Captions}}

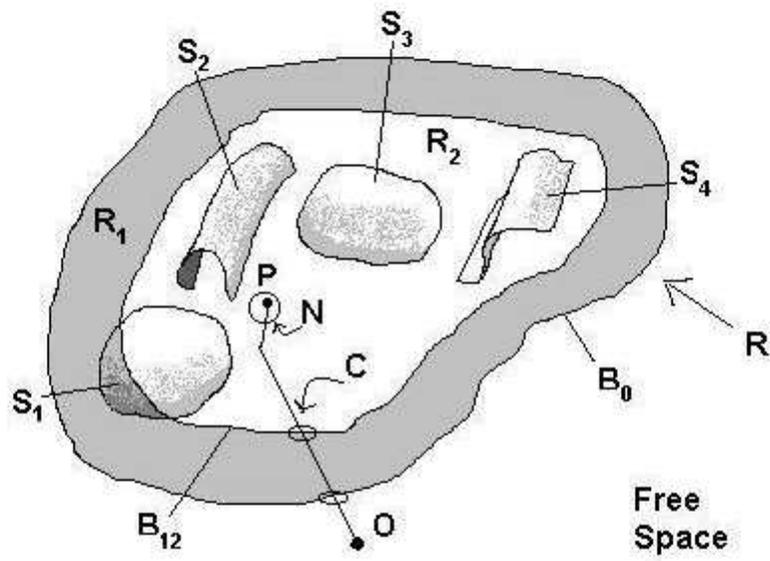
{\bf Fig.1}. An illustration of a simply connected region $R$ constituted in
this case by only two simply connected subregions $R_1$ and $R_2$ having a
boundary $B_{12}.$The boundary with free space is denoted by $B_0.$ The set
of non-nesting surfaces $S\ \ $have four elements $S_i$ , $i=1,..4.$ two of
them open and other two closed ones. A piece wise straight curve $C$ joining
any interior point $P$ of $R$ and a point $O$ in the free space is also
shown.\vspace{0.25in}$\vspace{0in}$

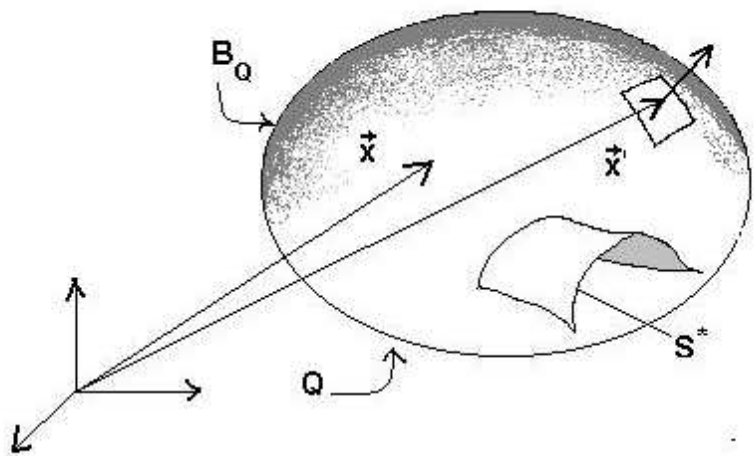
{\bf Fig.2. }Picture representing the region $Q$ in which a field $\varphi $
satisfies the Laplace equation and its value at the point $\overrightarrow{x}
$ is given by the Green integral (5).\vspace{0.25in}

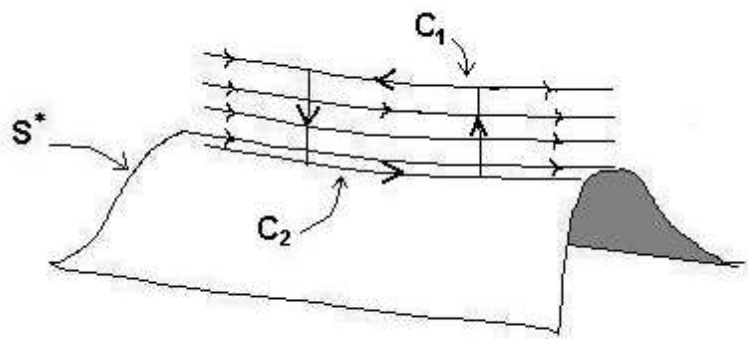
{\bf Fig.3. }The contour employed in the line integral in Eq. (6).\vspace{%
0.25in}

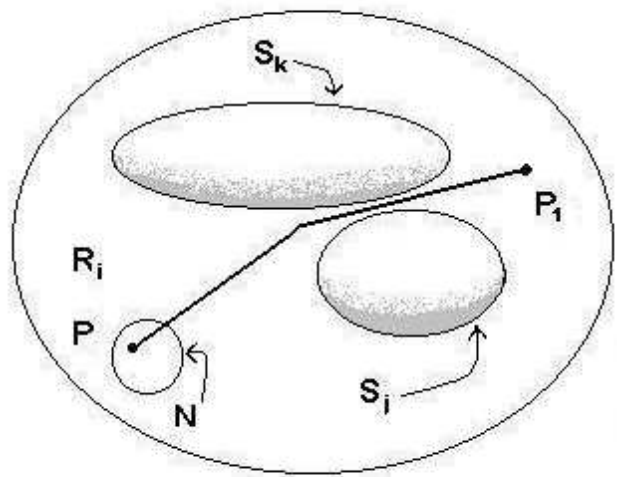
{\bf Fig.4. }Picture of the region{\bf \ }$R_i\ $and the open neighborhood $%
N $ in which the field $\varphi $ vanish exactly . A piece wise straight
line curve $C\ $joining a point $P\in N\ $and certain point $P_1$ in $R_i$
is also shown.\vspace{0.25in}

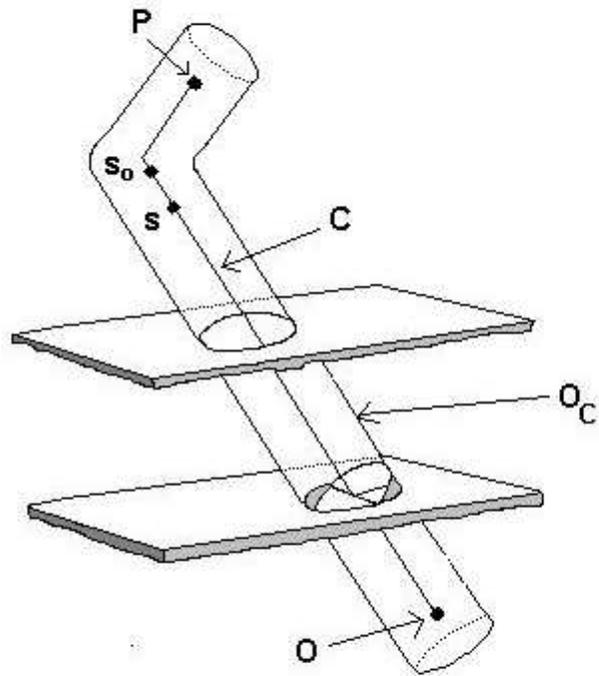
{\bf Fig.5. }Scheme of the curve $C$ and the open region$\ O_C$ containing
it.

\end{document}